\documentstyle[aasms4,12pt]{article}
\begin{document}

\title{A NEW SUPERWIND WOLF-RAYET GALAXY MRK 1259}
\author{Youichi Ohyama$^1$, Yoshiaki Taniguchi$^{1, 2}$, and Roberto
Terlevich$^2$}

\vspace {1cm}

\affil{$^1$Astronomical Institute, Tohoku University, Aoba, Sendai 980-77,
Japan}
\affil{$^2$Royal Greenwich Observatory, Madingley Road, Cambridge, CB3 0EZ, UK}
\received{Nov. 8, 1996}
\accepted{Feb. 7, 1997} 
---------

\begin{abstract}
 
We report the discovery of a starburst-driven wind
(superwind) from the starburst nucleus galaxy
Mrk 1259. The estimated number ratio of  Wolf-Rayet (WR) to O stars
amounts to $\simeq$ 0.09.
While the nuclear emission-line region is due to usual photoionization by
massive stars, 
the circumnuclear emission-line regions show anomalous line ratios that can be
due to cooling shocks.
Since the host galaxy seems to be a face-on disk galaxy and the excitation
conditions of the
circumnuclear emission-line regions show the spatial symmetry, we consider
that we are seeing
the superwind nearly from a pole-on view.
Cooling  shock models may  explain the observed emission line ratios
of the circumnuclear regions although
a factor of 2 overabundance of nitrogen is necessary.
All these suggest that  the high-mass enhanced
starburst occurred $\sim 5 \times 10^6$ years ago in the nuclear region of
Mrk 1259.

\end{abstract}
 
---------
 

--------

\section{INTRODUCTION}

The starburst-driven winds 
(hereafter superwinds: Heckman, Armus, \& Miley 1990)
may play an important role in chemical and 
dynamical evolution of galaxies (Chevalier \& Clegg 1985;
Tomisaka \& Ikeuchi 1988; Taniguchi et al. 1988;
  Heckman et al. 1990; Lehnert \& Heckman 1996).
It is expected that they would  occur at the epoch of galaxy formation if
the intense starbursts are associated with the galaxy formation (cf. Arimoto
\& Yoshii 1987).
In order to understand the basic nature of these winds, any detailed,
systematic study of
starburst galaxies would be helpful.
We have conducted our observing program to study nearby starburst galaxies 
based on very high S/N, long-slit optical 
spectroscopy at the Okayama Astrophysical Observatory.
During the course of this program, we have found a new superwind galaxy, 
Mrk 1259 (=IC 630 = PGC 31636),
which is one of the brightest nearby starburst galaxies 
(Balzano 1983; Keel, de Grijp, \& Miley 1988)
at a distance of 33.5 Mpc [the recession velocity with respect to
the Galactic Standard of Rest, $V_{\rm GSR} = 2,513$ km s$^{-1}$
(de Vaucouleurs et al. 1991) with a
Hubble constant $H_0$ = 75 km s$^{-1}$ Mpc $^{-1}$].

---------------
 
\section{OBSERVATIONS}
 
The spectroscopic observations were made at Okayama Astrophysical Observatory, 
National Astronomical Observatory of Japan in March 29, 1993.
The New Cassegrain Spectrograph was attached to the Cassegrain focus of the
188 cm reflector.
A Photometrics 516 $\times$ 516 CCD with pixel size of 20 $\mu$m $\times$ 20
$\mu$m was used, 
giving a spatial resolution of 1.46 arcsec per pixel by 1 $\times$ 2 binning.
The observations were carried out under photometric conditions.
A 1.8 arcsec slit with a length of 300 arcsec was used with a grating of 150 
grooves mm$^{-1}$ blazed at 5000 \AA, giving a spectral resolution 
(FWHM) of 9 \AA\ at 5000 \AA.
The position angle was set to 90 degree.
The wavelength coverage was set to 4400 - 6900 \AA\ in order to include WR
features around 4650 \AA.
Six spectra were taken in succession. Since the integration time of each
spectrum is 1,800 sec,
the total one is 10,800 sec.
These spectra are combined with median mode to improve the signal to noise
ratio. The final spectroscopic resolution is 11 \AA.
Sky regions are interactively set at the both far sides of the nucleus and
subtracted by 
means of the first-order polynomial fitting.
The flux scale 
is calibrated by using a standard star (HD 84937) which was observed just 
before the integration of the first exposure.
The seeing size during the observations was about 3 arcsec.


\section{RESULTS}


\subsection{\it The nuclear spectrum and the WR feature}

The nuclear spectrum (the central 2.9 $\times$ 1.5 arcsec$^2$ region) 
is shown in Fig. 1.
We also show the enlarged spectrum in order to see the WR bump around 4650\AA.
Although Keel, de Grijp, \& Miley (1988) reported the detection 
of the WR feature in Mrk 1259,
any detailed description was not given. Our spectroscopy has confirmed 
the presence of the WR feature.
The WR bump is evident only in the central region ($<$ 3 arcsec) 
as is observed for some starburst
galaxies (Sugai \& Taniguchi 1992).
Although the so-called WR feature consists of many emission lines 
(e.g., Sargent \& Filippenko 1991; 
Vacca \& Conti 1992), our spectrum shows that the  
broad HeII$\lambda$4686, NIII$\lambda$4638, and NV$\lambda$4619 
emission lines are dominant in Mrk 1259.
Although NV$\lambda$4619 is rare in WR stars,
this emission is also detected in the SBm galaxy NGC 4214
(Sargent \& Filippenko 1991).
Since the strength of the NIII$\lambda$4638 is nearly comparable 
with that of HeII$\lambda$4686, 
the dominant subtype of WN stars is considered to be late WN (WNL) 
(see, e.g., Conti \& Massey 1989).
However, the equivalent width of NV+NIII features at $\lambda$4604-4640
is usually weaker than that of HeII in WN stars (Smith, Shara, \& Moffat 1996).
Therefore, the assignment of WNL may be relatively uncertain.
We do not find firm evidence for CIV$\lambda$4658 feature which originates 
from carbon WR (WC) stars.
The emission line data, corrected for the reddening $E(B-V)$ = 0.58 mag,
 are summarized in Table 1.
The dereddened [OIII]$\lambda$5007/H$\beta$ ratio 
gives an oxygen abundance of $N$(O)/$N$(H) = 0.42 [$N$(O)/$N$(H)]$_\odot$
based on the empirical relation derived by Edmunds \& Pagel (1984).


We estimate the number of WNL stars and the number ratio
of WNL to OV stars.
The dereddened HeII$\lambda$4686 emission luminosity
is estimated to be 1.1$\times 10^{40}$ erg s$^{-1}$
the central 470 pc $\times$ 240 pc region.
Adopting the average HeII luminosity of a WNL star,
1.7$\times 10^{36}$ erg s$^{-1}$ (Vacca \& Conti 1992), we obtain
the number of WNL stars, $N$(WNL) $\simeq$ 6500.
The dereddened flux ratio of HeII$\lambda$4686 to H$\beta$ emission,
$I$(HeII)/$I$(H$\beta$) = 0.085, 
provides $N$(WNL)/$N$(OV) $\simeq 0.09$
by using the new method proposed by Schaerer (1996: see also Vacca 1994).
In order to estimate the age of the starburst, we compare the relationship
between $I$(HeII$\lambda$4686)/$I$(H$\beta$)  and the 
equivalent width of H$\beta$ emission, EW(H$\beta$) 
with the starburst model of Schaerer (1996) in Fig. 2.
Since the oxygen abundance in Mrk 1259 is 0.42 times of the solar value,
we adopt the model for $Z=0.008$ in Schaerer (1996)
where $Z$ is the total metal abundance in mass. The model result
is kindly provided by D. Schaerer for this study.
The model is an instantaneous starburst model with Salpeter's 
initial mass function together with the lower and higher mass cutoffs 
of 0.8 $M_\odot$ and 120 $M_\odot$, respectively.
The observational data point drops on a point of the model locus,
implying that the starburst is almost instantaneous (i.e., the duration
$< 1 \times 10^6$ years) and 
the age of the starburst is $\simeq 5.5 \times 10^6$ years.

-----

\subsection{\it The spatial variation of emission-line ratios}

Our spectrum shows that the emission-line regions are 
spatially extended up to 18 arcsec from
the nucleus ($\simeq$ 2.9 kpc).
The observed forbidden emission lines are [OIII]$\lambda$4959,5007,  
[NII]$\lambda$6583, [SII]$\lambda\lambda$6717, 6731, and
[OI]$\lambda$6300\footnote{ 
We abbreviate the forbidden emission lines as follows if not explicitly
indicated: 
[NII] = [NII]$\lambda$6583, [OIII] = [OIII]$\lambda$5007; [OI] =
[OI]$\lambda$6300, 
and [SII] = [SII]$\lambda\lambda$6717, 6731.}.
Following Veilleux \& Osterbrock (1987), we investigate the excitation
conditions of the 
extended emission-line regions in Fig. 3.
Surprisingly, we find that the outer emission-line regions show the AGN-like
excitation
while the nuclear region can be interpreted by the usual photoionization by
massive stars.


We examine if these excitation properties are real
because it is possible that the underlying 
stellar Balmer absorption may dilute the Balmer emission
in some cases (see Taniguchi et al. 1996 and references therein).
In the case of Mrk 1259, although 
there is  a trace of the underlying Balmer absorption (H$\beta$)
in the nuclear spectrum, there is no such feature in the 
outer region spectra.
If a large number of A-type stars would reside in the circumnuclear region,
the largest possible equivalent width of Balmer absorption may be
EW(H$\alpha$) $\simeq$ 5.6 \AA~ and EW(H$\beta$) $\simeq$ 7.9 \AA~ for a 
star cluster with age of 5 $\times 10^8$ years (Bica \& Alloin 1986).
Even if we apply these absorption corrections for our data,
the AGN-like excitation properties still remain for the circumnuclear regions.
It is thus   concluded that the excitation properties shown in Fig. 3 are real.


\section{DISCUSSION}

\subsection{\it Heating sources of the outer AGN-like emitting regions}

We discuss the heating source of the outer AGN-like emitting regions.
First, we examine a possibility of photoionization by massive stars.
A serious problem for the photoionization model is the observed large 
[OI]/H$\alpha$ ratio (see Filippenko \& Terlevich 1992; Shields 1992).
This ratio is enhanced in higher density regions in a partly ionized region
because [OI] has a large critical density
($N_{\rm cr} \simeq 1.8 \times 10^6$ cm$^{-3}$: cf. De Robertis \&
Osterbrock 1984).
However, the [SII] doublet ratio in our spectrum shows that 
the electron density is $\sim 10^3$ cm$^{-3}$ for the nuclear region and it
decreases 
to only  $\sim 10^2$ cm$^{-3}$ at radius of 10 arcsec. 
This trend is typical for such nuclear starburst galaxies 
(Sugai \& Taniguchi 1992; Lehnert \& Heckman 1996).
Even if we presume the presence of clumpy dense regions, 
we cannot explain the enhancement in both
[OI]/H$\alpha$ and [SII]/H$\alpha$ simultaneously. 
Another possibility may be the effect of 
unusually large population of WR stars in Mrk 1259
because the effective temperatures of WR stars are higher than those of
usual OB stars
and thus much more hard energy photons are available for the excitation
(Terlevich \& Melnick 1985; see, however, Schmutz, Leitherer, \&
Gruenward 1992).
Since, however, the WR stars are located in the central region, 
the ionization parameter should decrease with
radius monotonically given the above density distribution.
Therefore, we cannot explain the higher [OI]/H$\alpha$ ratio and the higher
[OIII]/H$\beta$ ratio in more distant regions simultaneously (cf. Boer,
Schulz, \& Keel  1992).
Accordingly, we may conclude that 
the simple photoionization model by massive stars cannot explain the 
observed spatial variation consistently.

An alternative heating source is cooling shocks 
(Daltabuit \& Cox 1972; Daltabuit, MacAlpine, \& Cox 1978; Heckman 1980;
Terlevich et al. 1992; Dopita \& Sutherland 1995).
It is natural to consider that strong shocks occur around the starburst nucleus
because  supernova explosions follow shortly  after
the starburst.
In fact, such shock-heated regions are observed in several starburst galaxies
(Taniguchi 1987; Heckman, Armus, \& Miley 1987; McCarthy, van Bruegel, \&
Heckman 1987; 
Heckman et al. 1990, 1995; Boer et al. 1992;
see also Kennicutt, Keel, \& Blaha 1989).
Recently Dopita \& Sutherland (1995) presented the optical diagnostic
diagrams based on
models of fast shocks in the low-density and magnetized medium.
Their diagnostic diagrams show that the AGN-like emission-line ratios are
explained 
by the cooling shocks whose velocities exceed a few 100 km s$^{-1}$.
In Fig. 3, we plot their model results for shock velocities ranging from 200
km s$^{-1}$
to 500 km s$^{-1}$ under the conditions
of $N_{\rm e} = 1$ cm$^{-3}$ and zero magnetic field.
The observed electron densities in Mrk 1259 ($\sim 10^{2-3}$ cm$^{-3}$) are
higher than
the model value. 
These models, however, explain the spatial variations 
of the emission line ratios basically
though there is a significant displacement between the models and the
observations 
in the [OIII]/H$\beta$ vs. [NII]/H$\alpha$ diagram.
These comparisons suggest that the outer regions are heated by faster shocks
than 
the inner regions. 
If the starburst occurred almost instantaneously (see section 3. 1.)
and the star formation rate decreased with time rapidly, it is expected that 
the gas in the outer region would be excited by 
faster shock because this shock is attributed to
the superwind blown out from the nuclear starburst region in an earlier phase 
(i.e., the phase of the most numerous supernova events). 
On the other hand, in the near nuclear region, 
the shock velocity would slow down because of the rapid cease of the wind.
In this way, the unusual spatial variations of 
the emission-line ratios can be understood by
{\it the slowing-down superwind scenario}.

Here we consider why there is the systematic difference only in 
the [OIII]/H$\beta$ vs. [NII]/H$\alpha$ diagram between the models and the
observations.
In order to investigate this difference more clearly, 
we show the following two diagrams in
Fig. 4; 1) [OI]/H$\alpha$ vs. [NII]/H$\alpha$, and 2) [SII]/H$\alpha$ vs.
[NII]/H$\alpha$.
Although these two diagrams show the systematic difference again, 
the difference seems due mostly to the offset in the [NII]/H$\alpha$ ratio;
i.e., the observed [NII]/H$\alpha$ ratios are systematically higher by a
factor of
2 than the model results.
Therefore, we consider that the difference may 
be attributed to the overabundance in nitrogen
with a factor of 2.
Since Mrk 1259 has a large number of WN stars, it is likely that
nitrogen-rich gas 
around the WN stars would be blown by the successive supernova events.
Such the nitrogen overabundance has also been reported 
in the starburst regions in NGC 5253 (Walsh \& Roy 1989).
It is interesting to note that Seyfert and LINER nuclei also show 
the nitrogen overabundance (Storchi-Bergmann \& Pastoriza 1989;
Storchi-Bergmann, Bica,  \& Pastoriza 1990; Storchi-Bergmann 1991).


\subsection{\it The superwind of Mrk 1259}

We consider the geometrical configuration of the superwind of Mrk 1259.
The superwind would be collimated by the circumnuclear disk gas,
and thus it tends to expand along the minor axis of the galactic disk 
just like the case of the archetypical starburst galaxy M82 (Bland \& Tully
1988).
Since the host galaxy of Mrk 1259 looks a nearly face-on disk galaxy,
it is considered that 
{\it we observe the superwind of Mrk 1259 from a nearly pole-on view}.
This is also advocated by the fact that the spatial variations of the
emission-line ratios
are almost symmetrical between the eastern and the western directions (see
Figs. 3 and 4).
After the superwind would come out from the circumnuclear region, it will
expand 
with a certain opening angle which would be determined by the initial
collimation
(Tomisaka \& Ikeuchi 1988).
Since it is considered that the expansion toward the galactic disk plane
may be more difficult than toward the galactic poles,
the linear length of the superwind along the line of sight
may be at least as long as the projected
spatial extent, 2.9 kpc. It is thus suggested
 that the full opening angle of the superwind
is less than 90$^\circ$.
The total spatial extent is estimated to be $\simeq 4$ kpc or longer.
Since the age of the starburst is $5.5 \times 10^6$ years, a mean wind
velocity is 
estimated to be $\simeq 710$ km s$^{-1}$.

The present observations show that 
the superwind of Mrk 1259 is observed from a nearly pole-on view.
Also, the spatial variations of the emission-line ratios suggest that
the superwind has been slowing down for these several 10$^6$ years.
Mrk 1259 will provide a unique chance to
investigate the  dynamical evolution of superwinds.

\vspace {0.5cm}

We would like to thank the staff of the Okayama Astrophysical Observatory
for kind
assistance of the observations.
We would like to thank Daniel Schaerer for providing us his 
starburst model for $Z=0.008$.
We thank Bill Vacca, Daniel Schaerer, Guillermo Tenorio-Tagle, and Brent Tully
 for useful discussion and comments.
YT thanks Don Hall, Len Cowie,  and Dave Sanders at Institute for Astronomy,
University of Hawaii, and Keith Tritton, Roberto \& Elena Terlevich, Isabel
Salamanca,
Itziar Aretxaga 
at Royal Greenwich Observatory for their warm hospitality.
This work was partly supported by the Ministry of Education, Science, and
Culture 
[No. 05452016 (FY 1994 and 1995)].
YO is a JSPS Research Fellow.


\begin{table}
\caption{Emission line data\tablenotemark{a}}
\begin{tabular}{ccc}
& \\
\tableline
\tableline
Line & $F/F$(H$\beta$)\tablenotemark{a} & $I/I$(H$\beta$)\tablenotemark{b} \\
\tableline
NV$\lambda$ 4619        & 0.030 & 0.035 \\
NIII$\lambda$4638       & 0.090 & 0.102 \\
HeII$\lambda$4686       & 0.078 & 0.085 \\
H$\beta$                & 1.00   & 1.00  \\
$[$OIII$]$$\lambda$5007 & 1.65   & 1.53  \\
HeI$\lambda$5876        & 0.41   & 0.28  \\
$[$OI$]$$\lambda$6300   & 0.093 & 0.056 \\
H$\alpha$               & 5.08   & 2.86  \\
$[$NII$]$$\lambda$6583  & 1.77   & 0.99 \\
HeI$\lambda$6678        & 0.065 & 0.036 \\
$[$SII$]$$\lambda$6717  & 0.40   & 0.22  \\
$[$SII$]$$\lambda$6731  & 0.41   & 0.22  \\
& & \\
$I$(H$\beta$) & \multicolumn{2}{c}{9.7$\times 10^{-13}$ 
ergs s$^{-1}$ cm$^{-2}$} \\
EW(H$\beta$) & \multicolumn{2}{c}{29.4 \AA} \\
$E_{\rm B-V}$ & \multicolumn{2}{c}{0.58 mag} \\
\tableline
\tablenotetext{a}{Observed emission-line ratio relative to H$\beta$.}
\tablenotetext{b}{Dereddened emission-line ratio relative to H$\beta$.}
\end{tabular}
\end{table}


\newpage
 
 
\figcaption{
The nuclear spectrum of Mrk 1259.
The two wavelength regions around $\lambda$5410 \AA~ and $\lambda$5740 \AA~ are 
contaminated by strong night sky lines (denoted by ^^ ^^ S").
The atmospheric absorption around 6810 \AA~ is denoted by ^^ ^^ A".
The inset spectrum  shows the WR features around 4650 \AA.
\label{fig1}}

\figcaption{
Diagram of $I$(HeII)/$I$(H$\beta$) vs. EW(H$\beta$). 
The observed data are compared with the starburst model provided by Daniel
Schaerer.
The right ordinate shows the age of the starburst (the red line).
\label{fig2}}
 
\figcaption{
The spatial variations of the emission-line ratios.
The shock model by Dopita \& Sutherland (1995) are also shown by plus marks
(green).
The dashed thick line in each panel shows the distinction between AGN and
HII regions
taken from Veilleux \& Osterbrock (1987).
The blue and red points correspond to the western (W) and the eastern (E)
parts from the nucleus (Nuc.), respectively.
Each spatial increment is 2.9 arcsec.
\label{fig3}}

\figcaption{
The spatial variations of the emission-line ratios:
a) log [OI]/H$\alpha$ vs. log [NII]/H$\alpha$ ({\it upper panel}), and 
b) log [SII]/H$\alpha$ vs. log [NII]/H$\alpha$ ({\it lower panel}).
The right panels show the close up of the nuclear region data.
The shock model by Dopita \& Sutherland (1995) are also shown by plus marks
(green).
The blue and red points correspond to the western and the eastern parts from
the nucleus, respectively.
Each spatial increment is 2.9 arcsec.
\label{fig4}}

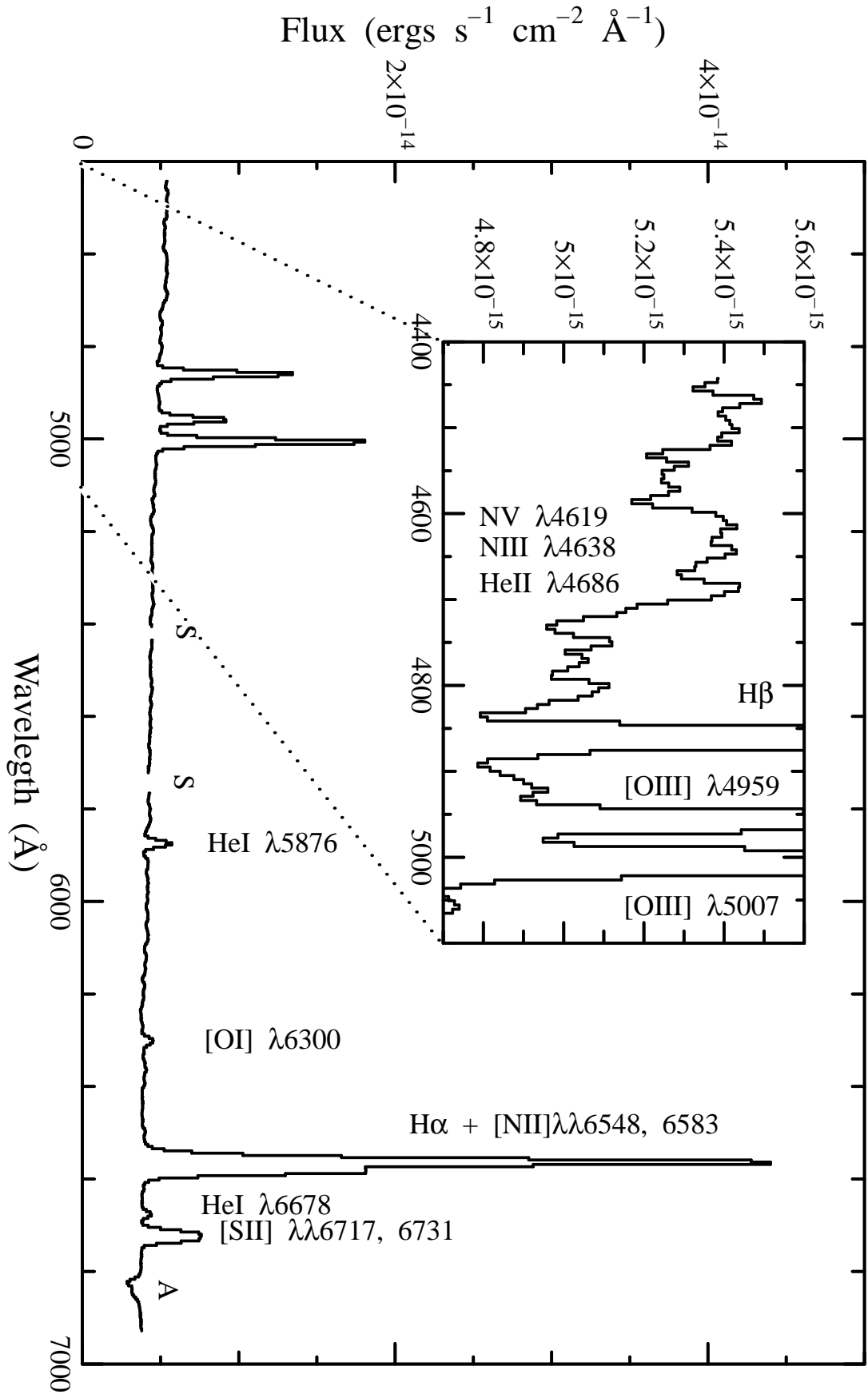
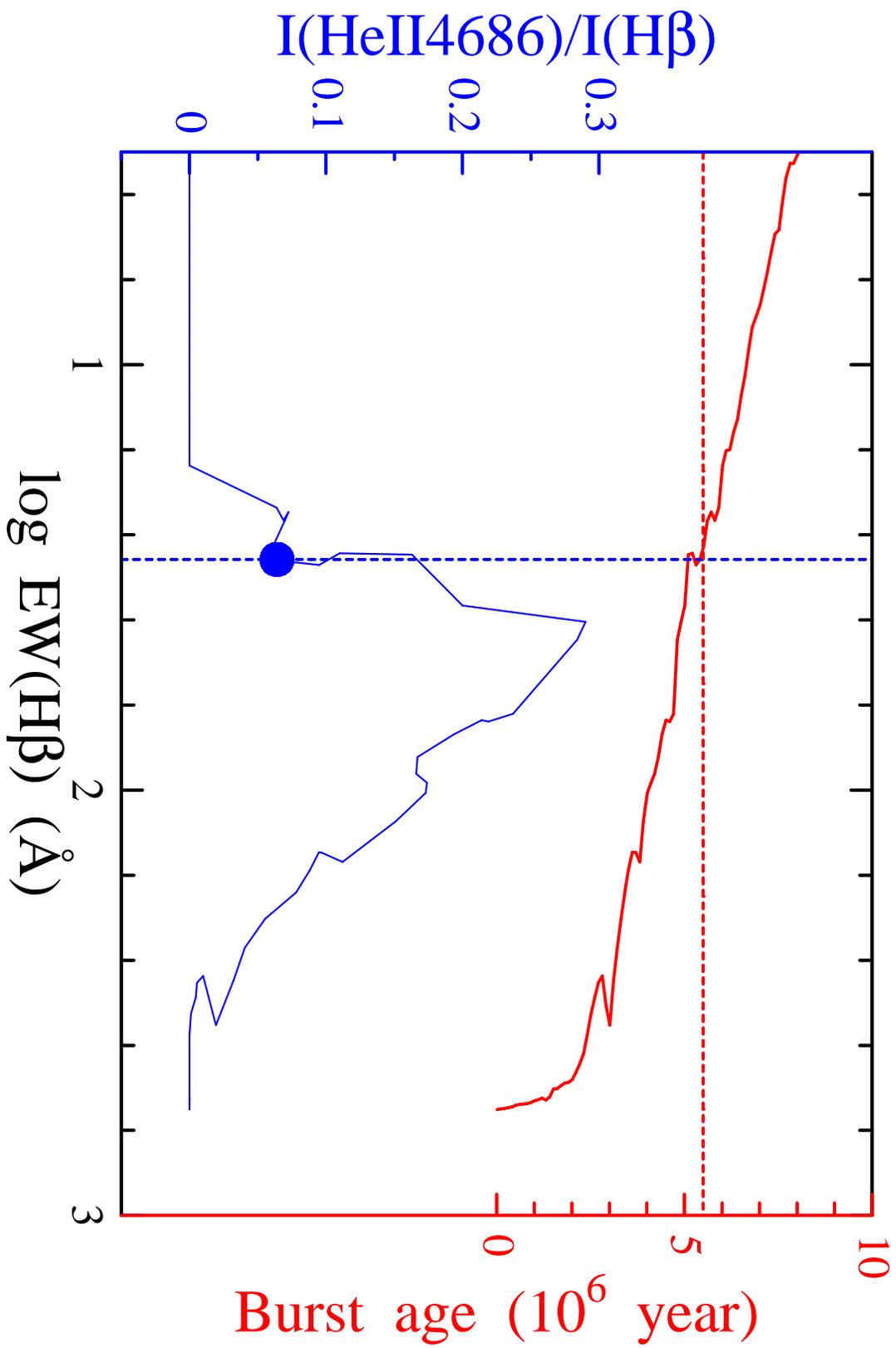
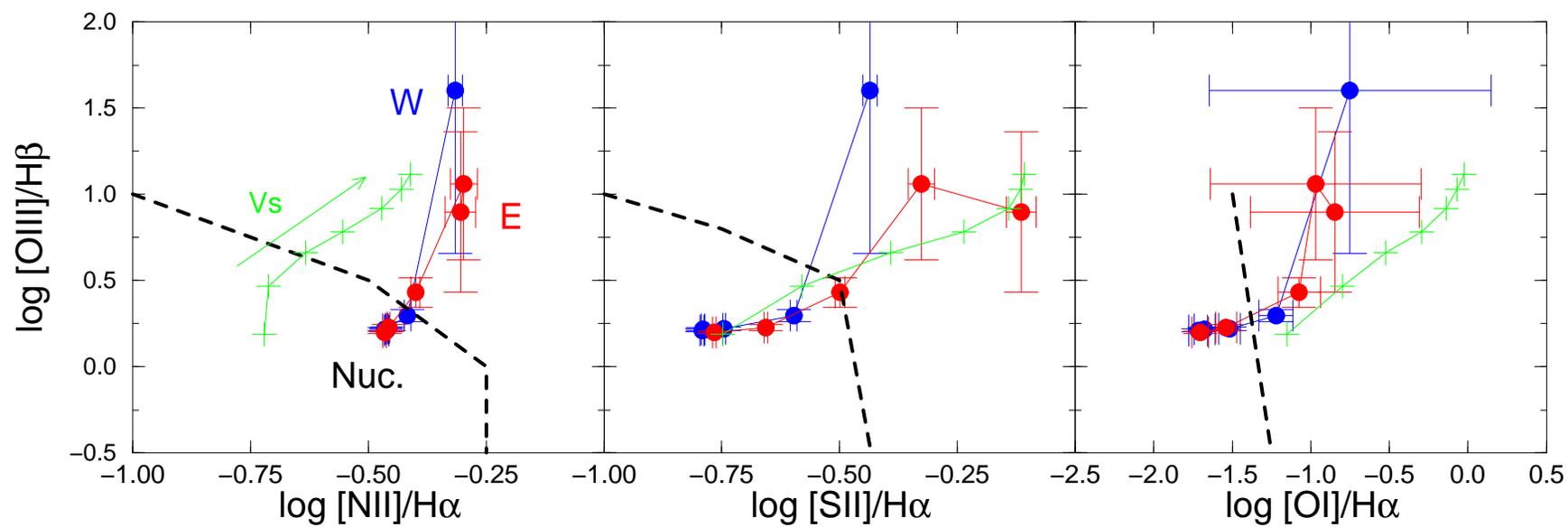
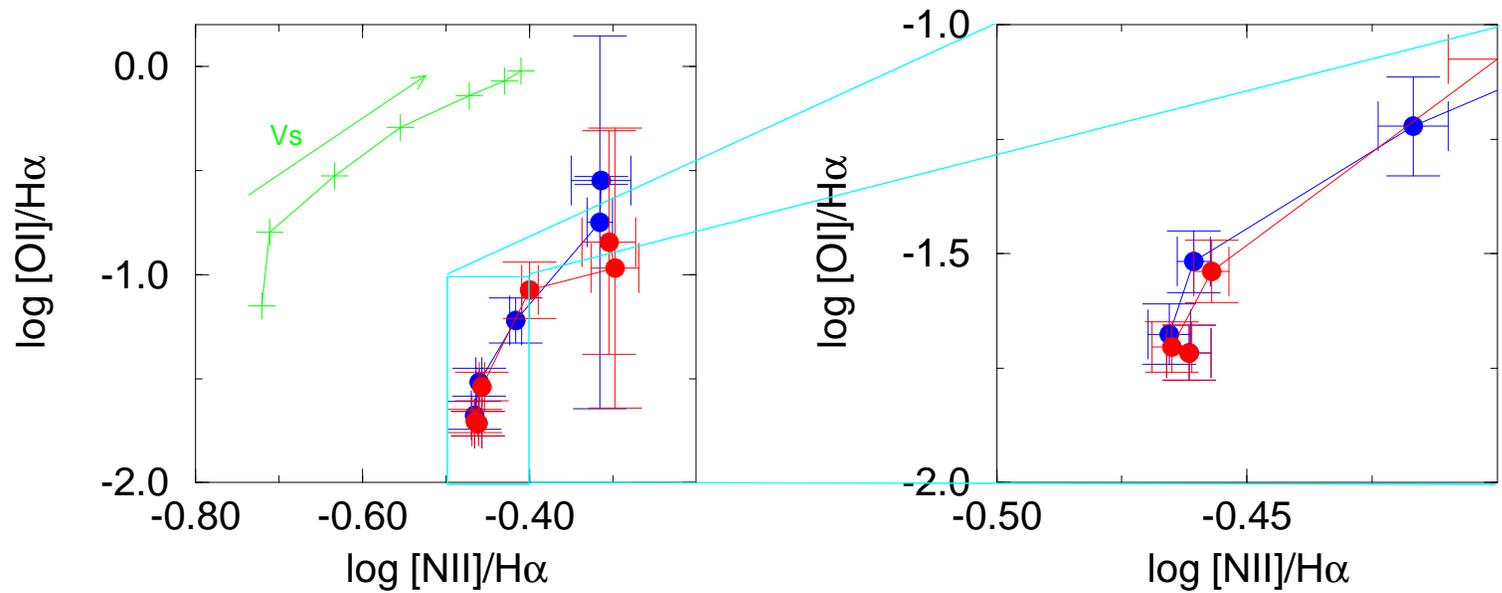
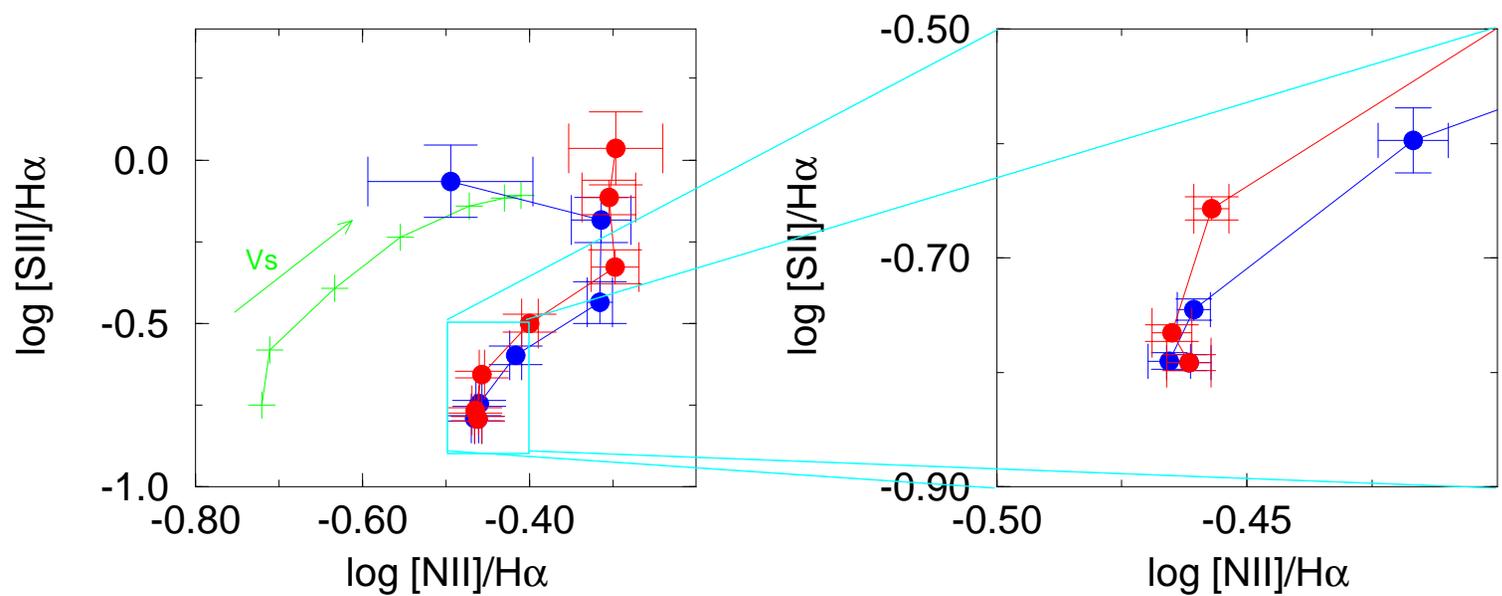
\end{document}